\newcounter{myctr}
\def\myitem{\refstepcounter{myctr}\bibfont\noindent\ifnum\themyctr>9\else\phantom{0}\fi\hangindent17pt\themyctr.\enskip}

\documentclass{ws-ijqi}
\usepackage{color}
\newcommand{\bra}[1]{\langle#1|}
\newcommand{\ket}[1]{|#1\rangle}

\begin{document}

\markboth{B. Bellomo, G. Compagno, A. D'Arrigo, G. Falci, R. Lo Franco, E. Paladino}
{Loss of nonlocal correlations due to adiabatic and quantum noise}

\catchline{}{}{}{}{}

\title{\textbf{DECAY OF NONLOCALITY DUE TO ADIABATIC AND QUANTUM NOISE IN THE SOLID STATE}}

\author{B. BELLOMO$^1$, G. COMPAGNO$^1$, A. D'ARRIGO$^2$, G. FALCI$^2$, R. LO FRANCO$^1$ and E. PALADINO$^2$}

\address{$^1$CNISM \& Dipartimento di Scienze Fisiche ed Astronomiche, Universit\`a di Palermo,
via Archirafi 36, 90123 Palermo, Italy.\\
lofranco@fisica.unipa.it}

\address{$^2$Dipartimento di Metodologie Fisiche e Chimiche, Universit\`a di Catania,
viale A. Doria 6, 95125 Catania, Italy \&  CNR - IMM MATIS\\
epaladino@dmfci.unict.it}

\maketitle

\begin{history}
\received{Day Month Year}
\revised{Day Month Year}
\end{history}

\begin{abstract}
We study the decay of quantum nonlocality, identified by the violation of the Clauser-Horne-Shimony-Holt (CHSH) Bell inequality, for two noninteracting Josephson qubits subject to independent baths with broadband spectra typical
of solid state nanodevices. The bath noise can be separated in an adiabatic (low-frequency) and in a quantum (high-frequency) part. We point out the qualitative different effects on quantum nonlocal correlations induced by adiabatic and quantum noise. A quantitaive analysis is performed for typical noise figures in Josephson systems. Finally we compare, for this system, the dynamics of nonlocal correlations and of entanglement.
\end{abstract}


\section{Introduction}	
The presence of quantum correlations in composite nanosystems is an essential resource for
quantum information  processing~\cite{Nielsen}.
Amongst the most relevant aspects of quantum correlations, of both  fundamental and applicative
role, are: entanglement~\cite{horodecki2009RMP}, quantum discord~\cite{ollivier2001PRL} and
nonlocality~\cite{Nielsen,bellclauser}. Studying these quantities for realistic quantum systems that are promising candidates for realizing a quantum computer is an interesting issue. Considerable development has been recently made towards the implementation of a solid-state quantum computer. In particular, superconducting high-fidelity~\cite{single-super,highfid} single qubit gates with coherence times of $\sim 1\mu$s are now available~\cite{schreier,vion}. Two-qubit logic gates have been proved in different laboratories~\cite{coupled-exp} and Bell states have been prepared up to $75\%$ of fidelity~\cite{bellgeneration}. In a circuit quantum electrodynamic framework, highly entangled two-qubit states with concurrence up to $94\%$ have been also generated, allowing the first implementation of basic quantum algorithms with a superconducting quantum processor\cite{dicarlo}.

In order to process quantum information, it is important to establish how long the relevant quantum
correlations can be maintained in noisy nanocircuits. Solid state noise may indeed
represent a serious limitation towards this goal. Josephson junction-based experimental setups are
often influenced by broadband and structured noise, whose typical power spectra show a $1/f$ low-frequency
behavior up to some cut-off frequency followed by a white or ohmic behavior~\cite{ithier,nak-spectrum}.
The presence of slow components in the environment makes the decay of the coherent signal strongly
dependent on the experimental protocol being used~\cite{vion,falci2005PRL}. Measurements protocols requiring
numerous repetitions are particularly sensitive to the unstable device calibration due to low-frequency
fluctuations. The main effect is dephasing due to defocusing of the measured signal.
Incoherent energy exchanges between system and environment, leading to relaxation and decoherence, occur
at typical operating frequencies (about $10$ GHz) where noise is white or ohmic.

Low- and  high-frequency noise affects quite differently single-qubit gates: adiabatic
(low-frequency) noise typically leads to algebraic decay, while quantum (high-frequency) noise to
exponential behavior~\cite{ithier,falci2005PRL}.
It is therefore important to evaluate the effect on quantum correlations time evolution
of adiabatic noise and  of its interplay with quantum noise. This analysis has been recently addressed in
detail for entanglement~\cite{pa-ct2010PRA}. Here we extend this study to the aspect of nonlocality.

Nonlocal correlations, i.e. correlations not reproducible in the framework
of Bell inequality tests by any classical local model,
are particularly important for quantum cryptography purposes~\cite{acin2006PRL}. In this paper, we consider
two noninteracting qubits subject to independent environments with typical broadband spectra. We use the
Clauser-Horne-Shimony-Holt (CHSH) inequality and
the maximum of the related Bell function~\cite{bellclauser,horodecki1995PLA}
to identify nonlocal correlations for classes of entangled initial states currently obtainable in
laboratory. In particular, we study the sensitivity of the time evolution of nonlocal correlations to the initial state purity and degree of entanglement.

\section{Model}
We consider a system composed by two identical superconducting qubits, namely $A$ and $B$,
each interacting with independent baths characterized by a broadband spectrum.
The total Hamiltonian is  $H_\textrm{tot}=H_A+H_B$, where each single-qubit Hamiltonian is
given by ($\hbar=1$)\cite{falci2005PRL,pa-ct2010PRA}
\begin{equation}\label{Hamiltonian}
H_\alpha = H_{Q, \alpha}-\hat{\xi}_\alpha \sigma_{z,\alpha}/2,\quad
H_{Q, \alpha} = -\vec{\Omega}_\alpha \cdot \vec{\sigma}_\alpha/2,
\end{equation}
where $ H_{Q, \alpha}$ refers to the qubit $\alpha=A,B$, $\vec{\sigma}_\alpha$ is the Pauli matrices vector,
$|\vec \Omega_\alpha|\equiv \Omega_\alpha$ the qubit frequency splitting and $\hat{\xi}_\alpha$ are collective environmental
variables  whose power spectra are $1/f$ at $f\in[\gamma_m,\gamma_M]$ (low-frequency noise)
and white or ohmic at frequencies of the order of the qubit splittings (high-frequency noise).
According to a standard model, noise with $1/f$ spectrum can be originated by an ensamble
of bistable fluctuators (BFs)~\cite{weissman1988RMP}. The physical origin of the fluctuators
depends on the specific setup. For instance, charge based devices
are extremely sensitive to background charge
fluctuations~\cite{falci2005PRL,paladino2002PRL,paladino2003ASSP,galperin06}.
Noise at higher frequencies instead may either originate from quantum
impurities~\cite{nak-spectrum}, possibly of the same physical origin, or being due to the circuitry.

Effects of the different parts of the spectrum can be treated in a multi-stage approach
introduced in Ref.\cite{falci2005PRL}.
Effects of low- and high-frequency components of the noise are distinguished
by decomposing $\hat{\xi}_\alpha \to \xi_\alpha(t)+\hat{\xi}_{f,\alpha}$~\cite{paladino2009}.
Stochastic variables $\xi_\alpha(t)$ describe low-frequency $1/f$ noise, and can be treated in
the adiabatic and longitudinal approximation. High-frequency
($\omega \sim \Omega_\alpha$) fluctuations $\hat{\xi}_{f,\alpha}$
are modeled by a Markovian bath  mainly leading to
spontaneous decay. Therefore, populations relax  due to quantum noise
($T_1$-type times), which also leads to secular dephasing
($T_2= 2 \, T_1$-type). Low-frequency noise provides
a defocusing mechanism determining further coherences decay.
In the Hamiltonian of Eq.~(\ref{Hamiltonian}) both the operating point
(angle $\theta_\alpha$ between $z$ and $\vec{\Omega}_\alpha$) and the qubit splitting $\Omega_\alpha$
are tunable. In the following we will consider both qubit operating at
the optimal working point, $\theta_\alpha=\pi/2$, where partial reduction of defocusing is
achieved\cite{vion,ithier}. In addition we will focus on the case of identical qubits
$\Omega_\alpha \equiv \Omega$.

The two-qubit density matrix elements will be evaluated in the computational basis
$\mathcal{B}=\{\ket{1}\equiv\ket{11},\ket{2}\equiv\ket{01}, \ket{3}\equiv\ket{10}, \ket{4}\equiv\ket{00}\}$,
where for each qubit we have $H_{Q,\alpha}\ket{0}=-\frac{\Omega}{2}\ket{0}$,
$H_{Q,\alpha}\ket{1}=\frac{\Omega}{2}\ket{1}$. Each subsystem ``qubit+environment'' evolves independently
so that, once known the single-qubit dynamics\cite{falci2005PRL}, the evolved two-qubit density matrix
will be readily obtained by a procedure reported in Refs.\cite{bellomo2007PRL,bellomo2008PRAprocedure}.

\section{Maximum of the Bell function \label{CHSHinequality}}
In this section we report the expression of the maximum of the Bell function for
the class of two-qubit states whose density matrix $\hat{\rho}_X$, in the standard computational
basis $\mathcal{B}$, has a ``X structure'', i.e.
\begin{equation}\label{Xstatesdensitymatrix}
   \hat{\rho}_X = \left(
\begin{array}{cccc}
  \rho_{11} & 0 & 0 & \rho_{14}  \\
  0 & \rho_{22} & \rho_{23} & 0 \\
  0 & \rho_{23}^* & \rho_{33} & 0 \\
  \rho_{14}^* & 0 & 0 & \rho_{44} \\
\end{array}
\right).
\end{equation}
This class of states is general enough to include the most common two-qubit states,
like Bell states (pure two-qubit maximally entangled states) and Werner states (mixture of Bell
states)~\cite{Nielsen,horodecki2009RMP,bellomo2008PRAprocedure}. Such a X form density matrix
may arise in a variety of physical situations \cite{expXstates}. A further remarkable feature of
X states is that, under various kinds of dynamics, their structure is maintained during
time evolution\cite{bellomo2007PRL,bellomo2008PRAprocedure}.

Using the Horodecki criterion\cite{horodecki1995PLA,bellomo2010PLA}, the maximum of the Bell
function can be expressed in terms of three functions
$u_1$, $u_2$ and $u_3$ of the density matrix elements as $B=2\sqrt{\mathrm{max}_{j>k}\{u_j+u_k\}}$,
where $j,k=1,2,3$. The CHSH inequality reads like $B\leq 2$, so that no classical local models are admitted
for states such that $B$ is larger than the classical threshold 2.
The three functions $u_j$ are\cite{bellomo2010PLA,mazzolapalermo2010PRA}
\begin{equation}\label{uXstate}
u_1=4(|\rho_{14}|+|\rho_{23}|)^2,\ u_2=(\rho_{11}+\rho_{44}-\rho_{22}-\rho_{33})^2,\
u_3=4(|\rho_{14}|-|\rho_{23}|)^2.
\end{equation}
Being $u_1$ always larger than $u_3$, the maximum of the Bell function for X states
is
\begin{equation}\label{Bellfunction}
B=\mathrm{max}\{B_1, B_2\},\quad B_1=2\sqrt{u_1+u_2},\ B_2=2\sqrt{u_1+u_3}.
\end{equation}
This quantity has been already studied in dynamical contexts of independent qubits each coupled to a
bosonic reservoir (cavity) with Markovian\cite{miranowicz2004PLA} and non-Markovian\cite{bellomo2008PRABell} characteristics. In the following, we shall investigate the maximum of the Bell function, $B$,
for our system of two independent Josephson qubits each interacting with individual baths.

\section{Initial states}
We consider extended Werner-like (EWL) two-qubit initial states\cite{bellomo2008PRAprocedure}
\begin{equation}\label{EWLstates}
    \hat{\rho}^\Phi=r \ket{\Phi}\bra{\Phi}+(1-r)I_4/4,\quad
    \hat{\rho}^\Psi=r \ket{\Psi}\bra{\Psi}+(1-r)I_4/4,
\end{equation}
whose pure parts are the one/two-excitation Bell-like states $\ket{\Phi}=a\ket{01}+b\ket{10}$,
$\ket{\Psi}=a\ket{00}+b\ket{11}$, where $|a|^2+|b|^2=1$. The purity parameter $r$ quantifies the mixedness
and $a$ sets the degree of entanglement of the initial state. The density matrix of EWL states, in the
computational basis, is non-vanishing only along the diagonal and anti-diagonal (X form) and this
structure is maintained at $t>0$ in the system we are considering. Using concurrence\cite{wootters1998PRL}
 $C$ to quantify entanglement, one can also notice that the initial entanglement is equal for both the
 EWL states of Eq.~(\ref{EWLstates}) and reads
$C_\rho^{\Phi}(0)=C_\rho^{\Psi}(0)=2\mathrm{max}\{0,(|ab|+1/4)r-1/4\}$. Initial states are thus entangled
for $r>r^\ast=(1+4|ab|)^{-1}$.

Entangled states with purity $\approx 0.87$ and fidelity to ideal Bell states $\approx 0.90$ have been experimentally generated\cite{dicarlo}. These states may be approximately described as EWL states with $r_\mathrm{exp}\approx0.91$.

\section{Time behavior of the maximum of the Bell function}
In this section we shall analyze the dynamics of the maximum of the Bell function, $B$ of
Eq.~(\ref{Bellfunction}), initially considering the case when only adiabatic noise is present and
in a second stage including the effect of quantum noise. We shall also compare the dynamics of $B$ and
of the concurrence $C$.

\subsection{Adiabatic noise}
The effect of low-frequency noise components is obtained by the Hamiltonian Eq.~(\ref{Hamiltonian})
 with $\hat{\xi}\approx\xi(t)$ treated in the adiabatic and longitudinal approximation, for which
single-qubit populations do not evolve in time\cite{falci2005PRL,pa-ct2010PRA}. The main effect of
low-frequency fluctuations is dephasing due to defocusing processes.
The leading order effect depends only on the noise variance, which can be estimated by independent
measurements of the amplitude  of the $1/f$ power spectrum on the uncoupled qubits, as follows
$S^{1/f}(\omega) =\pi \Sigma^2 [\ln(\gamma_M/\gamma_m) \, \omega]^{-1}$.
By exploiting the single-qubit coherences
determined in this case\cite{falci2005PRL}, we can construct the two-qubit density
matrix\cite{bellomo2007PRL,pa-ct2010PRA} at time $t$.
Choosing initial states of the form (\ref{EWLstates})  $B$ is obtained from  Eq.~(\ref{Bellfunction}).

We find that, under adiabatic noise, $B(t) \equiv B_\mathrm{ad}(t)$ is the same for both the initial EWL
states of Eq.~(\ref{EWLstates}).
The explicit expressions of
$B_\mathrm{ad}(t)$ and of the times when $B_\mathrm{ad}=2$, that is when a Bell inequality
``violation sudden death'' (VSD) occurs, are given by
\begin{equation}\label{Bellfunctionadiab-tVSDtimes}
B_\mathrm{ad}(t)=2r\sqrt{1+\frac{4|ab|^2\Omega^2}{\Omega^2+\Sigma ^4 t^2 }},\quad
t_\mathrm{VSD}^\mathrm{ad}=\frac{\Omega}{\Sigma^2}\sqrt{\frac{4|ab|^2r^2}{(1-r)^2}-1}.
\end{equation}
For any $r<1$, the  Bell inequality is violated at a finite time.
Instead, for pure initial entangled states ($r=1$), $B_{ad}(t)$ approaches asymptotically the classical threshold.
\begin{figure}
\begin{center}
{\includegraphics[width=0.45 \textwidth]{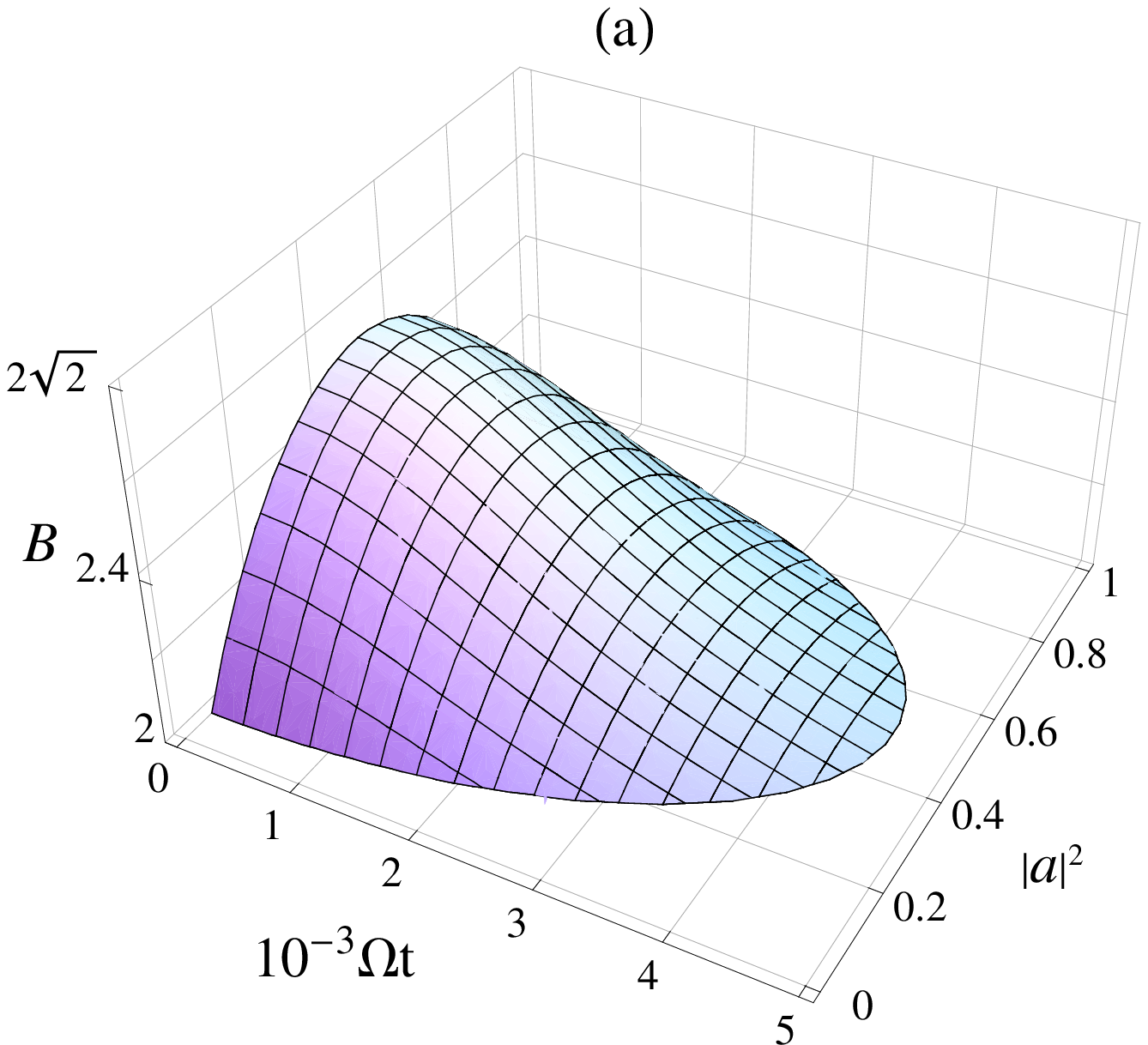}
\hspace{0.5 cm}
\includegraphics[width=0.45 \textwidth]{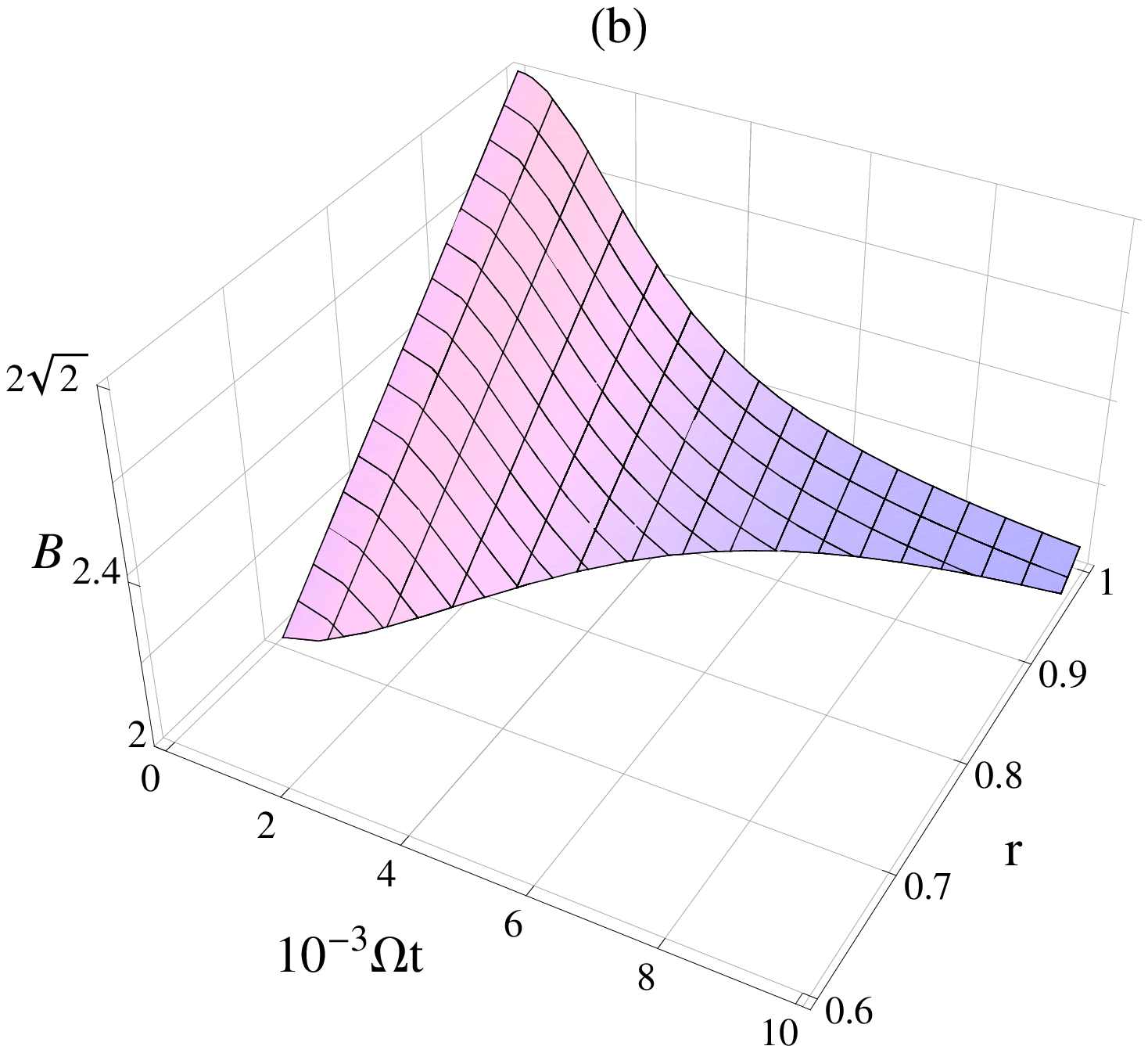}}
\caption{\label{fig:Bintime}\footnotesize
 Maximum of the Bell function, $B$, at $\theta=\pi/2$ and $\Sigma/\Omega=0.02$ under adiabatic
(low frequency) noise. Panel (a) $B$ as a function of dimensionless time $\Omega t$ and $|a|^2$
($r=0.9$); Panel (b) $B$ as a function of dimensionless time $\Omega t$ and $r$ ($a=1/\sqrt2$).}
\end{center}
\end{figure}
The evolution of $B_{ad}(t)$ is displayed in Fig.~\ref{fig:Bintime} for 
$\Sigma/\Omega=0.02$, which is a typical figure of $1/f$ noise in  single-qubit
experiments\cite{vion,ithier,falci2005PRL}. The dependence on the initial degree of entanglement, parameterized by $|a|^2$,
for fixed $r=0.9$, is reported in panel (a). Sensitivity to the initial purity $r$, for fixed
$a=1/\sqrt2$, is shown in panel (b). We note that the maintenance of nonlocal correlations ($B>2$) strongly depends on the
purity of the initial state $r$, while the dependence on $|a|^2$ is weaker and symmetric
around $|a|^2=1/2$. The ``violation sudden death'' of  $B_{ad}(t)$ for $r<1$ is also clearly visible.

\subsection{Interplay of adiabatic and quantum noise}
We now analyze the interplay of adiabatic and quantum noise simultaneously affecting the two units.

When quantum noise (high-frequency, $f\sim\Omega$) is included, it adds up to the defocusing channel
leading to both extra exponential decay of the coherences and evolution of the populations.
This is due to the fact that, in practical situations, decay rates are much less sensitive than
phases to fluctuations of control parameters.
Single-qubit populations are obtained by the Born-Markov master equation.
They decay with a relaxation rate $T_1^{-1}=S_f(\Omega)/2$ to
asymptotic values which depend in general on $\Omega$ and on temperature $T$\cite{pa-ct2010PRA} .
For typical values of $\Omega$ ($\sim10^{11}$ rad/s) and T ($\sim0.04$ K),
asymptotic values of the excited, $|4\rangle$, and of the ground state, $|1\rangle$,
populations are practically zero and one, respectively. On the other hand, the coherences acquire
an additional exponential decaying
factor\cite{falci2005PRL,pa-ct2010PRA}. Using the reported single-qubit density matrix elements
for this case\cite{falci2005PRL}, we construct the new two-qubit density
matrix\cite{bellomo2007PRL,pa-ct2010PRA} from which we determine $B(t)$.
\begin{figure}
\begin{center}
{\includegraphics[width=0.75 \textwidth]{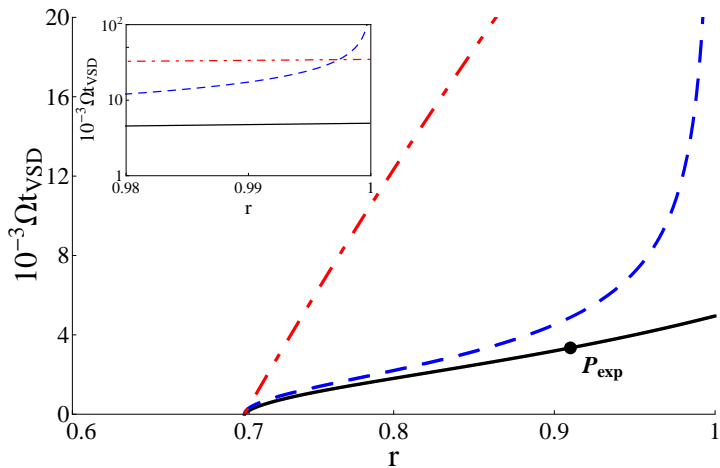}}
\caption{\label{fig:tVSD}\footnotesize Dependence of the VSD time on the purity $r$ ($a=1/\sqrt{2}$)
for initial state  $\hat{\rho}^\Phi$. The behavior for $\hat{\rho}^\Psi$ is also quantitatively similar.
The blue dashed curve is $\Omega t_\mathrm{VSD}^\mathrm{ad}$, red dot-dashed curve is for quantum noise,
black curve is the result of adiabatic and quantum noise altogether. Noise parameters are
$\Sigma=0.02\Omega$, $S_f(\omega)=2\times10^{6}$ s$^{-1}$.
In addition, $\Omega=10^{11}$ rad/s, $\theta=\pi/2$, $T=0.04$ K. The inset zooms the region where $r\approx1$.
The point $P_\mathrm{exp}$ corresponds to $r_\mathrm{exp}\approx0.91$ where $\Omega t_\mathrm{VSD}\approx 3350$.}
\end{center}
\end{figure}
The maximum of the Bell functions for the two initial EWL states are now formally nonequivalent,
this is a qualitative difference with the adiabatic noise case.
However, for the typical experimental parameters involved in
such nanosystems, quantitatively they do not differ significantly (their difference being always
$\lesssim10^{-3}$ in the violation region). We are interested to the VSD times $t_\mathrm{VSD}$ when $B=2$.
These times are plotted in Fig.~\ref{fig:tVSD} as a function of the purity  $r$ for fixed
$a=1/\sqrt{2}$ (maximally entangled pure part), with noise parameters having values retrieved in
experiments\cite{vion,ithier}. In particular, the white noise level is $S_f=2\times 10^6$s$^{-1}$.
We distinguish the cases of only adiabatic noise, only quantum noise and their interplay.
For $r<1$, adiabatic noise suppresses nonlocal correlations on a much shorter time scale than quantum noise.
For high purity levels, the VSD time due to quantum noise is instead shorter than the adiabatic VSD time (inset of Fig.~\ref{fig:tVSD}), which goes to infinity for pure states (see Eq.~(\ref{Bellfunctionadiab-tVSDtimes})). However, a quantitative estimate of the amount of nonlocal correlations preserved before $B=2$ indicates that, for typical
amplitudes of $1/f$ and white noise, adiabatic noise considerably reduces the amount of nonlocal correlations on a short time scale even for $r\to 1$. A finite value of $t_\mathrm{VSD}$ is ensured for {\em any} initial state only because of
quantum noise.

\subsection{Maximum of the Bell function versus concurrence}
\begin{figure}
\begin{center}
\includegraphics[width=0.66 \textwidth]{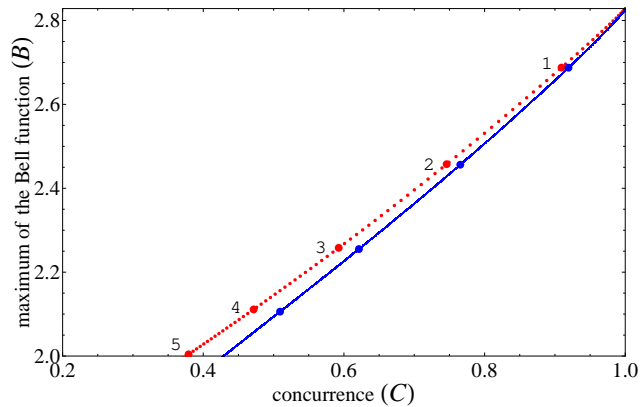}
\caption{\label{fig:BvsCintime}\footnotesize Maximum of the Bell function $B$ versus concurrence
$C$ for the two initial Bell states $\ket{\Phi}$ (blue solid curve) and $\ket{\Psi}$ (red dotted curve)
in the presence of  both adiabatic and quantum noise.
The points on the curves labeled with $i=1,2,\ldots,5$ indicates
the values of $(C,B)$ at  times $10^{-3}\Omega t=i$. $B$ for $\ket{\Phi}$ decays a little
bit faster than that for $\ket{\Psi}$. Noise parameters as in Fig.~\ref{fig:tVSD}.}
\end{center}
\end{figure}
Here we compare the dynamics of the maximum of the Bell function $B$ with the evolution of the concurrence
$C$, so to establish their connection in such a solid-state system. We consider initial preparation in the two Bell states  $\ket{\Phi}=(\ket{01}+\ket{10})/\sqrt{2}$ and $\ket{\Psi}=(\ket{00}+\ket{11})/\sqrt{2}$. The behavior for different initial values of $a$ and $r$ does not differ qualitatively. Remarkably, for the considered system there is  a one-to-one correspondence between $B$ and $C$ during the dynamics for both initial states, as shown in Fig.~\ref{fig:BvsCintime}. This property has already been observed in other physical contexts\cite{mazzolapalermo2010PRA,bellomo2008PRABell} (atomic qubits in cavities). However, the common behavior for both initial states is not predictable a priori. In general,
it may strongly depend on the physical system and on the specific initial state~\cite{mazzolapalermo2010PRA}.

The slightly different time dependence for the two Bell states is evidenced by the different distance along the curves of the dots pointing to times $10^{-3}\Omega t=i$ ($i=1,\ldots,5$). In addition it is clearly visible the  threshold value of $C$ below which there is no violation anymore. Starting from the initial state $\ket{\Phi}$ ($\ket{\Psi}$) we find that for
$C\leq0.43$ ($C\leq0.38$) the maximum of the Bell function $B\leq2$, so that we cannot be sure of the presence of nonlocal correlations in this region.

\section{Conclusions}
In this paper we have investigated the time evolution of nonlocal correlations (nonlocality),
identified by the maximum of the  Bell function $B$ when it violates the CHSH-Bell inequality
($B>2$), between two noninteracting Josephson qubits subject to independent baths
with broadband noise typical of solid state nanosystems.
In particular, an adiabatic (low-frequency) noise and a quantum (high-frequency) noise can be
distinguished. We have shown that, while adiabatic noise has the main effect on nonlocality decay,
it is the quantum noise that induces a complete disappearance of quantum nonlocal correlation for
any initial state (even for pure maximally entangled states).
In particular, we have also reported the times when $B=2$ ($t_\mathrm{VSD}$), after which there is no more certainty of the presence of quantum nonlocal correlations. We have finally compared, for this system, the dynamics of nonlocal correlations and that of entanglement, quantified by the
concurrence $C$. We have found that a one-to-one correspondence between $B$ and $C$ occurs in time,
independently on the form of the initial two-qubit state. Moreover, we obtained thresholds values of
$C$ below which the Bell inequality is not violated ($B\leq2$).

The results presented in this paper provide new insights towards the possibility to exploit
nonlocal quantum correlations for quantum information processing with superconducting nanocircuits.

\end{document}